\documentclass{article}
\usepackage{spconf,amsmath,graphicx,amsfonts,epsfig}
\usepackage{times}
\usepackage{amsbsy}
\usepackage{latexsym}
\usepackage{amssymb}

\title{Robust Auxiliary Vector Filtering with Constrained Constant Modulus Design for Beamforming \vspace{-0.5em}}

\name{Lei Wang and Rodrigo C. de Lamare \vspace{-0.75em}}
\address{Communications Research Group, Department of Electronics, University of York, YO10 5DD, UK.\\Email:\{lw517,rcdl500\}@ohm.york.ac.uk \vspace{-0.5em}}

\begin{document}
\ninept \maketitle \linespread{1.2}
\begin{abstract}
This paper proposes an auxiliary vector filtering (AVF) algorithm
based on a constrained constant modulus (CCM) design for robust
adaptive beamforming. This scheme provides an efficient way to deal
with filters with a large number of elements. The proposed
beamformer decomposes the adaptive filter into a constrained
(reference vector filters) and an unconstrained (auxiliary vector
filters) components. The weight vector is iterated by subtracting
the scaling auxiliary vector from the reference vector. The scalar
factor and the auxiliary vector depend on each other and are jointly
calculated according to the CCM criterion. The proposed robust AVF
algorithm provides an iterative exchange of information between the
scalar factor and the auxiliary vector and thus leads to a fast
convergence and an improved steady-state performance over the
existing techniques. Simulations are performed to show the
performance and the robustness of the proposed scheme and algorithm
in several scenarios.
\end{abstract}
\section{Introduction}
Adaptive beamforming techniques are widely used in numerous
applications such as radar, wireless communications, and sonar
\cite{Trees}-\cite{Jian}, to detect or improve the reception of a
desired signal while suppressing interference at the output of a
sensor array. Currently, the beamformers designed according to the
constrained minimum variance (CMV) and the constrained constant
modulus (CCM) criteria are among the most used criteria due to their
simplicity and effectiveness. The CMV criterion aims to minimize the
beamformer output power while maintaining the array response on the
direction of the desired signal. The CCM criterion is a positive
measure (Chapter 6 in \cite{Jian}) of the deviation of the
beamformer output from a constant modulus (CM) condition subject to
a constraint on the array response of the desired signal. Compared
with the CMV, the CCM-based beamformers exhibit superior performance
in many severe scenarios (e.g., steering vector mismatch) since the
positive measure provides more information for parameter estimation
with constant modulus signals.

For the design of adaptive beamformers, numerous adaptive filtering
algorithms have been developed using constrained optimization
techniques \cite{Haykin}, \cite{Lamare2}. The stochastic gradient
(SG) and recursive least squares (RLS) \cite{Haykin}, \cite{Lamare2}
are popular methods with different tradeoffs between performance and
complexity. A major drawback is that they require a large number of
samples to reach the steady-state when the array size is large. In
dynamic scenarios, filters with many elements usually provide a poor
performance in tracking signals embedded in interference and noise.
The multistage Wiener filter (MSWF) \cite{Goldstein} provides a way
out of this dilemma. The MSWF employs the minimum mean squared error
(MMSE) criterion and its extended versions with the CMV and the CCM
criteria are reported in \cite{Honig}, \cite{Lamare}. Another
cost-effective technique is the auxiliary vector filtering (AVF)
\cite{Pados} algorithm. In this scheme, an auxiliary vector is
calculated by maximizing the cross correlation between the outputs
of the reference vector filter and the previously auxiliary vector
filters. The weight vector is obtained by subtracting the scaling
auxiliary vector from the reference vector. In \cite{Pados2}, the
AVF algorithm iteratively generates a sequence of filters that
converge to the CMV filter with a small number of samples. Its
application in adaptive beamforming has been reported in
\cite{Mathews}.

Motivated by the fact that the CCM-based beamformers outperform the
CMV ones for the CM signals, we propose an AVF algorithm based on
the CCM design for robust adaptive beamforming. The beamformer
structure decomposes the adaptive filter into a constrained
(reference vector filters) and an unconstrained components
(auxiliary vector filters). The constrained component is initialized
with the array response of the desired signal to start the iteration
and to ensure the CCM constraint, and the auxiliary vector in the
unconstrained component can be iterated to meet the CM criterion.
The weight vector is computed by means of suppressing the scaling
unconstrained component from the constrained part. The main
difference from the existing AVF technique is that, in the proposed
CCM-based algorithm, the auxiliary vector and the scalar factor
depend on each other and are jointly calculated according to the CM
criterion (subject to different constraints). The proposed method
provides an iterative exchange of information between the auxiliary
vector and the scalar factor and also exploits the information about
the CM signals, which leads to an improved performance. Simulations
exhibit the robustness of the proposed method in typical scenarios
including array mismatches.

The rest of this paper is organized as follows: we outline a
system model and the problem statement in Section 2. The proposed
scheme is introduced and the CCM-AVF algorithm is developed in
Section 3. Simulation results are provided and discussed in
Section 4, and conclusions are drawn in Section 5.

\section{System Model and CCM Beamformer Design}
Consider $q$ narrowband signals that impinge on a uniform linear
array (ULA) of $m$ ($m\geq q$) sensor elements. The sources are
assumed to be in the far field with directions of arrival (DOAs)
$\theta_{0}$,\ldots,$\theta_{q-1}$. The $i$th received vector
$\boldsymbol x(i)\in\mathbb C^{m\times 1}$ can be modeled as
\begin{equation} \label{1}
\centering {\boldsymbol x}(i)={\boldsymbol {A}}({\boldsymbol
{\theta}}){\boldsymbol s}(i)+{\boldsymbol n}(i),~~~ i=1,\ldots,N,
\end{equation}
where
$\boldsymbol{\theta}=[\theta_{0},\ldots,\theta_{q-1}]^{T}\in\mathbb{R}^{q
\times 1}$ is the signal DOAs, ${\boldsymbol A}({\boldsymbol
{\theta}})=[{\boldsymbol a}(\theta_{0}),\ldots,{\boldsymbol
a}(\theta_{q-1})]\in\mathbb{C}^{m \times q}$ comprises the signal
steering vectors ${\boldsymbol a}(\theta_{k})=[1,e^{-2\pi
j\frac{d}{\lambda_{c}}cos{\theta_{k}}},\ldots$, $e^{-2\pi
j(m-1)\frac{d}{\lambda_{c}}cos{\theta_{k}}}]^{T}\in\mathbb{C}^{m
\times 1}$, $(k=0,\ldots,q-1)$, where $\lambda_{c}$ is the
wavelength and $d$ is the inter-element distance of the ULA
($d=\lambda_c/2$ in general), ${\boldsymbol s}(i)\in \mathbb
C^{q\times 1}$ is the source data, ${\boldsymbol n}(i)\in\mathbb
C^{m\times 1}$ is assumed to be a zero-mean spatially white Gaussian
process, $N$ is the number of snapshots, and $(\cdot)^{T}$\ stands
for transpose. To avoid mathematical ambiguities, the steering
vectors $\boldsymbol a(\theta_{k})$ are normalized and considered to
be linearly independents. The output of the beamformer is
\begin{equation} \label{2}
\centering y(i)={\boldsymbol w}^H(i) {\boldsymbol x}(i),
\end{equation}
where ${\boldsymbol w}(i)=[w_{1}(i),\ldots,w_{m}(i)]^{T}\in\mathbb
C^{m\times 1}$ is the complex weight vector of the beamformer, and
$(\cdot)^{H}$ stands for Hermitian transpose.

With the signals introduced in (\ref{1}) and (\ref{2}), we can
present the CCM beamformer design by minimizing the following cost
function
\begin{equation}\label{3}
J_{\textrm{cm}}\big(\boldsymbol w(i)\big)=\mathbb
E\big\{\big[|y(i)|^{2}-\nu\big]^{2}\big\},~~
\textrm{subject~to}~~{\boldsymbol w}^{H}(i){\boldsymbol
a}(\theta_{0})=1,
\end{equation}
where $\theta_0$ is the direction of the signal of interest (SOI)
and $\boldsymbol a(\theta_{0})$ denotes the corresponding steering
vector. The cost function is the expected deviation of the squared
modulus of the array output to a constant, say $\nu=1$. The
constraint is set to maintain the power of the SOI and to ensure the
convexity of the cost function. The weight expression obtained from
(\ref{3}) is
\begin{equation}\label{4}
\boldsymbol w(i+1)=\boldsymbol R^{-1}(i)\big\{\boldsymbol
p(i)-\frac{\big[\boldsymbol p^H(i)\boldsymbol R^{-1}(i)\boldsymbol
a(\theta_0)-1\big]\boldsymbol a(\theta_0)}{\boldsymbol
a^H(\theta_0)\boldsymbol R^{-1}(i)\boldsymbol a(\theta_0)}\big\},
\end{equation}
where $\boldsymbol R(i)=\mathbb E[|y(i)|^2\boldsymbol
x(i)\boldsymbol x^H(i)]\in\mathbb C^{m\times m}$, $\boldsymbol
p(i)=\mathbb E[y^{\ast}(i)\boldsymbol x(i)]\in\mathbb C^{m\times
1}$, and $(\cdot)^{\ast}$ denotes complex conjugate. Note that
(\ref{4}) is a function of previous values of $\boldsymbol w(i)$
(since $y(i)=\boldsymbol w^H(i)\boldsymbol x(i)$) and thus must be
initialized to start the iteration. We keep the time index $i$ in
$\boldsymbol R(i)$ and $\boldsymbol p(i)$ for the same reason. The
calculation of the weight vector is costly due to the matrix
inversion. The SG or RLS type algorithms can be employed to reduce
the computational load but suffer from a poor performance when the
dimension $m$ is large.

\section{Proposed CCM Beamformer Design\\ and AVF Algorithm}
In this section, we introduce a CCM-based adaptive filtering
structure for beamforming and develop an efficient CCM-AVF algorithm
for robust adaptive beamforming.
\subsection{Proposed CCM Beamformer}
We define the cost function for the beamformer design, which is
\begin{equation}\label{5}
J_{\textrm{av}}\big(\boldsymbol w(i)\big)=\mathbb
E\big\{\big[\boldsymbol w^H(i)\tilde{\boldsymbol
x}(i)-\nu\big]^2\big\},
\end{equation}
where $\tilde{\boldsymbol x}(i)=y^{\ast}(i)\boldsymbol x(i)$ can be
viewed as a new received vector to the beamformer and $\nu=1$ is set
in accordance with (\ref{3}).

To obtain the weight solution for the time index $i$, we start the
iteration by initializing the weight vector
\begin{equation}\label{6}
\boldsymbol w_0(i)=\boldsymbol a(\theta_0)/\|\boldsymbol
a(\theta_0)\|^2.
\end{equation}

Then, we subtract a scaling auxiliary vector (unconstrained
component) that is orthogonal to $\boldsymbol a(\theta_0)$ from
$\boldsymbol w_0(i)$ (constrained component) and obtain
\begin{equation}\label{7}
\boldsymbol w_1(i)=\boldsymbol w_0(i)-\mu_1(i)\boldsymbol g_1(i),
\end{equation}
where $\boldsymbol g_1(i)\in\mathbb C^{m\times1}$ with
$\boldsymbol g_1^H(i)\boldsymbol a(\theta_0)=0$, and $\mu_1(i)$ is
a scalar factor to control the weight of $\boldsymbol g_1(i)$. The
auxiliary vector is supposed to capture the signal components in
$\tilde{\boldsymbol x}(i)$ that are not from the direction
$\theta_0$. The aim of (\ref{7}) is to suppress the disturbance of
the unconstrained component while maintaining the contribution of
the SOI. The cost function in (\ref{5}) appears in unconstrained
form since the constraint has been incorporated in the weight
adaptation.

\subsection{Proposed CCM-AVF Algorithm }
From (\ref{7}), it is necessary to determine the auxiliary vector
$\boldsymbol g_1(i)$ and the scalar factor $\mu_1(i)$ for the
calculation of $\boldsymbol w_1(i)$. Assuming $\boldsymbol g_1(i)$
is known, $\mu_1(i)$ can be obtained by minimizing $\mathbb
E\{[\boldsymbol w_1(i)\tilde{\boldsymbol x}(i)-1]^2\}$.
Substituting (\ref{7}) into this cost function, computing the
gradient with respect to $\mu_1(i)$ and equating it to zero, we
obtain
\begin{equation}\label{8}
\mu_1(i)=\frac{\boldsymbol g_1^H(i)\tilde{\boldsymbol
R}(i)\boldsymbol w_0(i)-\boldsymbol g_1^H(i)\tilde{\boldsymbol
p}(i)}{\boldsymbol g_1^H(i)\tilde{\boldsymbol R}(i)\boldsymbol
g_1(i)},
\end{equation}
where $\tilde{\boldsymbol R}(i)=\mathbb E[\tilde{\boldsymbol
x}(i)\tilde{\boldsymbol x}^H(i)]\in\mathbb C^{m\times m}$ and
$\tilde{\boldsymbol p}(i)=\mathbb E[\tilde{\boldsymbol
x}(i)]\in\mathbb C^{m\times 1}$. Note that $\mu_1(i)=0$, i.e.,
$\tilde{\boldsymbol R}(i)\boldsymbol w_0(i)=\tilde{\boldsymbol
p}(i)$ needs to be avoided here since the design is equivalent to a
matched filter if it happens.

On the other hand, the calculation of the auxiliary vector
$\boldsymbol g_1(i)$ should take the conditions $\boldsymbol
g_1^H(i)\boldsymbol a(\theta_0)=0$ and $\boldsymbol
g_1^H(i)\boldsymbol g_1(i)=1$ into account. The constrained
minimization problem with respect to $\boldsymbol g_1(i)$ can be
transformed by the method of Lagrange multipliers into
\begin{equation}\label{9}
\begin{split}
J_{\textrm{L}}\big(\boldsymbol w_1(i)\big)=&\mathbb
E\big\{\big[\boldsymbol w_1^H(i)\tilde{\boldsymbol
x}(i)-1\big]^2\big\}\\
&-2~\mathfrak{R}\big\{\lambda_1\big[\boldsymbol g_1^H(i)\boldsymbol
g_1(i)-1\big]-\lambda_2\boldsymbol g_1^H(i)\boldsymbol
a(\theta_0)\big\},
\end{split}
\end{equation}
where $\lambda_1$ and $\lambda_2$ are scalar Lagrange multipliers.
For the sake of mathematical accuracy, we note that the cost
function to be minimized is phase invariant, namely, if $\boldsymbol
g_1(i)$ satisfies it, so does $\boldsymbol g_1(i)e^{j\phi}$ for any
phase $\phi$. To avoid any ambiguity, we assume that only one
auxiliary vector can be obtained.

Following the procedure to get $\mu_1(i)$, the auxiliary vector can
be expressed by
\begin{equation}\label{10}
\boldsymbol g_1(i)=\frac{\mu_1^{\ast}(i)\tilde{\boldsymbol
p}_{\textrm{y}}(i)-\lambda_2\boldsymbol a(\theta_0)}{\lambda_1},
\end{equation}
where $\tilde{\boldsymbol p}_{\textrm{y}}(i)=\mathbb
E\big[\big(1-\tilde{y}(i)\big)^{\ast}\tilde{\boldsymbol
x}(i)\big]\in\mathbb C^{m\times 1}$ and $\tilde{y}(i)=\boldsymbol
w^H(i)\tilde{\boldsymbol x}(i)$. We keep the time index $i$ in
$\tilde{\boldsymbol p}_{\textrm{y}}(i)$ since it is a function of
$\boldsymbol w(i)$, which must be initialized to provide an
estimation about $\tilde{y}(i)$ and to start the iteration.

The expression of $\boldsymbol g_1(i)$ is utilized to enforce the
constraints and solve for $\lambda_1$ and $\lambda_2$. Indeed, we
have
\begin{equation}\label{11}
\lambda_1=\Bigg\|\mu_1^{\ast}(i)\tilde{\boldsymbol
p}_{\textrm{y}}(i)-\frac{\mu_1^{\ast}(i)\boldsymbol
a^H(\theta_0)\tilde{\boldsymbol p}_{\textrm{y}}(i)}{\|\boldsymbol
a(\theta_0)\|^2}\boldsymbol a(\theta_0)\Bigg\|,
\end{equation}
\begin{equation}\label{12}
\lambda_2=\frac{\mu_1^{\ast}(i)\boldsymbol
a^H(\theta_0)\tilde{\boldsymbol p}_{\textrm{y}}(i)}{\|\boldsymbol
a(\theta_0)\|^2},
\end{equation}
where $\|\cdot\|$ denotes the Euclidean norm. Substitution of
$\lambda_1$ and $\lambda_2$ back in (\ref{10}) leads to $\boldsymbol
g_1(i)$ that satisfies the constraints and minimizes (with
$\mu_1(i)$) the squared deviation of $\tilde{y}(i)$ from the CM
condition, yielding
\begin{equation}\label{13}
\boldsymbol g_1(i)=\frac{\mu_1^{\ast}(i)\tilde{\boldsymbol
p}_{\textrm{y}}(i)-\frac{\mu_1^{\ast}(i)\boldsymbol
a^H(\theta_0)\tilde{\boldsymbol p}_{\textrm{y}}(i)}{\|\boldsymbol
a(\theta_0)\|^2}\boldsymbol
a(\theta_0)}{\big\|\mu_1^{\ast}(i)\tilde{\boldsymbol
p}_{\textrm{y}}(i)-\frac{\mu_1^{\ast}(i)\boldsymbol
a^H(\theta_0)\tilde{\boldsymbol p}_{\textrm{y}}(i)}{\|\boldsymbol
a(\theta_0)\|^2}\boldsymbol a(\theta_0)\big\|}.
\end{equation}

So far, we have detailed the first iteration of the proposed
CCM-AVF algorithm for time index $i$, i.e., $\boldsymbol w_0(i)$
in (\ref{6}), $\boldsymbol w_1(i)$ in (\ref{7}), $\mu_1(i)$ in
(\ref{8}), and $\boldsymbol g_1(i)$ in (\ref{13}), respectively.
In this procedure, $\tilde{\boldsymbol x}(i)$ can be viewed as a
new received vector that is processed by the adaptive filter
$\boldsymbol w_1(i)$ \big(first estimation of $\boldsymbol
w(i)$\big) to generate the output $\tilde{y}(i)$, in which,
$\boldsymbol w_1(i)$ is determined by minimizing the mean squared
error between the output and the desired CM condition. This
principle is suitable to the following iterations with
$\boldsymbol w_2(i), \boldsymbol w_3(i), \ldots$.

Now, we consider the iterations one step further and express the
adaptive filter as
\begin{equation}\label{14}
\boldsymbol w_2(i)=\boldsymbol
w_0(i)-\sum_{k=1}^{2}\mu_k(i)\boldsymbol g_k(i)=\boldsymbol
w_1(i)-\mu_2(i)\boldsymbol g_2(i),
\end{equation}
where $\mu_2(i)$ and $\boldsymbol g_2(i)$ will be calculated based
on the previously identified $\mu_1(i)$ and $\boldsymbol g_1(i)$.
$\mu_2(i)$ \big($\mu_2(i)\neq0$\big) is chosen to minimize the cost
function $\mathbb E\{[\boldsymbol w_2^H(i)\tilde{\boldsymbol
x}(i)-1]^2\}$ under the assumption that $\boldsymbol g_2(i)$ is
known beforehand. Thus, we have
\begin{equation}\label{15}
\mu_2(i)=\frac{\boldsymbol g_2^H(i)\tilde{\boldsymbol
R}(i)\boldsymbol w_1(i)-\boldsymbol g_2^H(i)\tilde{\boldsymbol
p}(i)}{\boldsymbol g_2^H(i)\tilde{\boldsymbol R}(i)\boldsymbol
g_2(i)},
\end{equation}
The auxiliary vector $\boldsymbol g_2(i)$ is calculated by the
minimization of the cost function subject to the constraints
$\boldsymbol g_2^H(i)\boldsymbol a(\theta_0)=0$ and\\ $\boldsymbol
g_2^H(i)\boldsymbol g_2(i)$$=1$, which is
\begin{equation}\label{16}
\boldsymbol g_2(i)=\frac{\mu_2^{\ast}(i)\tilde{\boldsymbol
p}_{\textrm{y}}(i)-\frac{\mu_2^{\ast}(i)\boldsymbol
a^H(\theta_0)\tilde{\boldsymbol p}_{\textrm{y}}(i)}{\|\boldsymbol
a(\theta_0)\|^2}\boldsymbol
a(\theta_0)}{\big\|\mu_2^{\ast}(i)\tilde{\boldsymbol
p}_{\textrm{y}}(i)-\frac{\mu_2^{\ast}(i)\boldsymbol
a^H(\theta_0)\tilde{\boldsymbol p}_{\textrm{y}}(i)}{\|\boldsymbol
a(\theta_0)\|^2}\boldsymbol a(\theta_0)\big\|}.
\end{equation}
The above iterative procedures are taken place at time index $i$
to generate a sequence of filters $\boldsymbol w_k(i)$ with $k=0,
1, \ldots$ being the iteration number. Generally, there exists a
maximum (or suitable) value of $k$, i.e., $k_{\textrm{max}}=K$,
that is determined by a certain rule to stop iterations and
achieve satisfactory performance. One simple rule, which is
adopted in the proposed CCM-AVF algorithm, is to terminate the
iteration if $\boldsymbol g_k(i)\cong\boldsymbol 0$ is achieved.
Alternative and more complicated selection rules can be found in
\cite{Mathews}. Until now, the weight solution at time index $i$
can be given by $\boldsymbol w(i)=\boldsymbol w_K(i)$. The
proposed CCM-AVF algorithm for the design of the CCM beamformer is
summarized in Table \ref{tab:AVF}.

\begin{table}[!t]
\centering
    \caption{PROPOSED CCM-AVF ALGORITHM}
    \label{tab:AVF}
    \begin{small}
        \begin{tabular}{l}
\hline
\bfseries {For the time index $i=1, 2, \ldots, N$}.\\
~~~\bfseries {Initialization:}\\
~~~~~~~~~~${\boldsymbol w}(i)=\boldsymbol w_0(i)=\frac{\boldsymbol a(\theta_0)}{\|\boldsymbol a(\theta_0)\|^2}$;~~~$\mu_0(i)=\textrm{small positive value}$.\\
~~~\bfseries {Iterative procedures:}\\
~~~~~~~~~~For $k=1, 2, \ldots, K$\\
~~~~~~~~~~~~~~$\boldsymbol
g_k(i)=\mu_{k-1}^{\ast}(i)\tilde{\boldsymbol
p}_{\textrm{y}}(i)-\frac{\mu_{k-1}^{\ast}(i)\boldsymbol
a^H(\theta_0)\tilde{\boldsymbol p}_{\textrm{y}}(i)}{\|\boldsymbol
a(\theta_0)\|^2}\boldsymbol a(\theta_0)$\\
~~~~~~~~~~~~~~\bfseries {if $\boldsymbol g_k(i)=\boldsymbol 0$ then EXIT}.\\
~~~~~~~~~~~~~~$\mu_k(i)=\frac{\boldsymbol g_k^H(i)\tilde{\boldsymbol
R}(i)\boldsymbol w_{k-1}(i)-\boldsymbol g_k^H(i)\tilde{\boldsymbol
p}(i)}{\boldsymbol g_k^H(i)\tilde{\boldsymbol R}(i)\boldsymbol
g_k(i)}$\\
~~~~~~~~~~~~~~$\boldsymbol w_k(i)=\boldsymbol
w_{k-1}(i)-\mu_{k}\boldsymbol
g_k(i)$\\
~~~\bfseries {Weight expression:}\\
~~~~~~~~~~~~~~$\boldsymbol w(i)=\boldsymbol w_K(i)$.\\
\hline
    \end{tabular}
    \end{small}
\end{table}

\subsection{Interpretations about Proposed CCM-AVF Algorithm}

There are several points we need to interpret in Table
\ref{tab:AVF}. First of all, initialization is important to the
realization of the proposed method. $\boldsymbol w(i)$ is set to
estimate $\tilde{y}(i)$ and so $\tilde{\boldsymbol R}(i)$,
$\tilde{\boldsymbol p}(i)$, and $\tilde{\boldsymbol
p}_{\textrm{y}}(i)$. $\boldsymbol w_0(i)$ is for the activation of
the weight adaptation. Note that the calculation of the scalar
factor, e.g., in (\ref{8}), is a function of $\boldsymbol g_1(i)$
and the auxiliary vector obtained from (\ref{13}) depends on
$\mu_1(i)$. It is necessary to initialize one of these quantities
to start the iteration. We usually set a small positive scalar
value $\mu_0(i)$ for simplicity. Under this condition, the
subscript of the scalar factor for the calculation of $\boldsymbol
g_k(i)$ should be replaced by $k-1$ instead of $k$, as shown in
Table \ref{tab:AVF}.

Second, the expected quantities $\tilde{\boldsymbol R}(i)$,
$\tilde{\boldsymbol p}(i)$, and $\tilde{\boldsymbol
p}_{\textrm{y}}(i)$ are not available in practice. We use a
sample-average approach to estimate them, i.e.,
\begin{equation}\label{17}
\begin{split}
&\hat{\tilde{\boldsymbol
R}}(i)=\frac{1}{i}\sum_{l=1}^i\tilde{\boldsymbol
x}(l)\tilde{\boldsymbol x}^H(l);~~~~~\hat{\tilde{\boldsymbol
p}}(i)=\frac{1}{i}\sum_{l=1}^{i}\tilde{\boldsymbol x}(l);\\
&\hat{\tilde{\boldsymbol
p}}_{\textrm{y}}(i)=\frac{1}{i}\sum_{l=1}^{i}\big(1-\tilde{y}(l)\big)^{\ast}\tilde{\boldsymbol
x}(i).
\end{split}
\end{equation}
where $\tilde{\boldsymbol R}(i)$, $\tilde{\boldsymbol p}(i)$, and
$\tilde{\boldsymbol p}_{\textrm{y}}(i)$ are substituted by their
estimates in the iterative procedure to generate $\boldsymbol
w_k(i)$. To improve the estimation accuracy, the quantities in
(\ref{17}) can be refreshed or further regularized during the
iterations. Specifically, we use $\boldsymbol w_k(i)$ in the
iteration step instead of $\boldsymbol w(i)$ in the initialization
to generate $y(i)$, and related $\tilde{\boldsymbol x}(i)$ and
$\tilde{y}(i)$, which are employed to update the estimates
$\hat{\tilde{\boldsymbol R}}(i)$, $\hat{\tilde{\boldsymbol
p}}(i)$, and $\hat{\tilde{\boldsymbol p}}_{\textrm{y}}(i)$.
Compared with $\boldsymbol w(i)=\boldsymbol
a(\theta_0)/\|\boldsymbol a(\theta_0)\|^2$, $\boldsymbol w_k(i)$
is more efficient to evaluate the desired signal. Thus, the
refreshment of the estimates based on the current $\boldsymbol
w_k(i)$ is valuable to calculate the subsequent scalar factor and
the auxiliary vector.

Third, we drop the normalization of the auxiliary vector
\cite{Pados2}-\cite{Qian}. Note that the calculated auxiliary
vectors $\boldsymbol g_k(i)$ are constrained to be orthogonal to
$\boldsymbol a(\theta_0)$. The orthogonality among the auxiliary
vectors is not imposed. Actually, the successive auxiliary vectors
do satisfy the orthogonality as verifies in our numerical results.
We omit the analysis about this characteristic here considering
the paper length.

The proposed CCM-AVF beamformer efficiently measures the expected
deviation of the beamformer output from the CM condition and provide
useful information for the proposed algorithm for dealing with
parameter estimation in many severe scenarios including low
signal-to-noise ratio (SNR) or steering vector mismatch. The
proposed CCM-AVF algorithm employs an iterative procedure to adjust
the weight vector for each time instant. The matrix inversion
appeared in (\ref{4}) is avoided and thus the computational cost is
limited. Since the scalar factor and the auxiliary vector depend on
each other, the proposed algorithm provides an iterative exchange of
information between them, which are jointly employed to update the
weight vector. This scheme leads to an improved convergence and the
steady-state performance that will be shown in the simulations.

\section{Simulations}

Simulations are performed for a ULA containing $m=40$ sensor
elements with half-wavelength interelement spacing. We compare the
proposed algorithm (CCM-AVF) with the SG \cite{Haykin}, RLS
\cite{Lamare2}, MSWF \cite{Lamare}, and AVF \cite{Pados2} methods.
With respect to each method, we consider the CMV and the CCM
criteria for beamforming. A total of $1000$ runs are used to get the
curves. In all experiments, BPSK sources' powers (desired user and
interferers) are $\sigma_{\textrm{S}}^2=\sigma_{\textrm{I}}^2=1$ and
the input SNR$=0$ dB with spatially and temporally white Gaussian
noise.

\begin{figure}[htb]
\begin{minipage}[h]{0.9\linewidth}
  \centering
  \centerline{\epsfig{figure=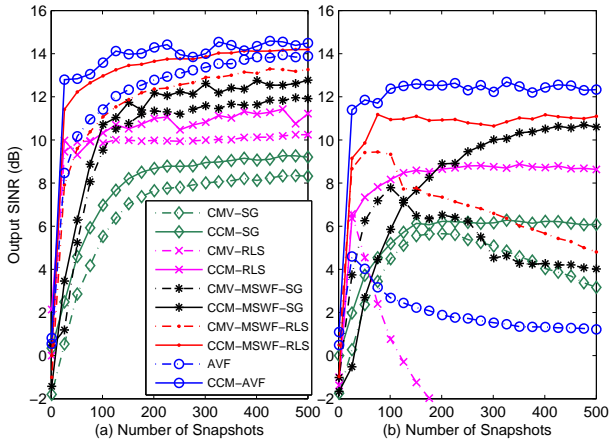,scale=0.57}} \vspace{-1em}\caption{Output SINR
versus the number of snapshots for (a) ideal steering vector; (b)
steering vector mismatch $1^o$.} \label{fig:AVF_sve}
\end{minipage}
\end{figure}

Fig. \ref{fig:AVF_sve} includes two experiments. There are $q=10$
users, including one desired user in the system. The scalar factor
is $\mu_0(i)=0.01$ and the iteration number is $K=3$. In
Fig.\ref{fig:AVF_sve} (a), the exact DOA of the SOI is known at the
receiver. All output SINR values increase to the steady-state as the
increase of the snapshots (time index). The RLS-type algorithms
enjoy faster convergence and better steady-state performance than
the SG-type methods. The proposed CCM-AVF algorithm converges
rapidly and reaches the steady-state with superior performance. The
CCM-based MSWF technique with the RLS implementation has comparative
fast convergence rate but the steady-state performance is not better
than the proposed method. In Fig. \ref{fig:AVF_sve} (b), we set the
DOA of the SOI estimated by the receiver to be $1^o$ away from the
actual direction. It indicates that the mismatch induces performance
degradation to all the analyzed algorithms. The CCM-based methods
are more robust to this scenario than the CMV-based ones. The
proposed CCM-AVF algorithm has faster convergence and better
steady-state performance than the other analyzed methods.

\begin{figure}[htb]
\begin{minipage}[h]{0.9\linewidth}
  \centering
  \centerline{\epsfig{figure=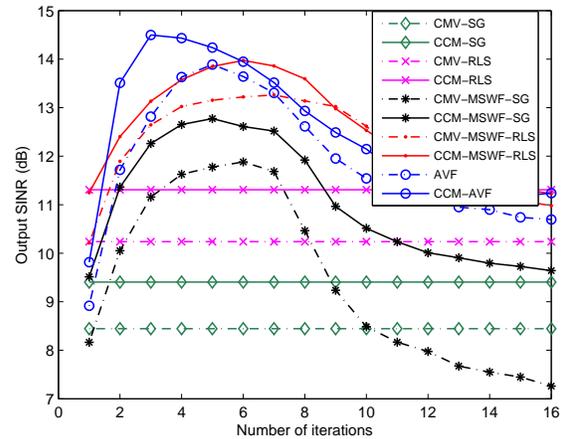,scale=0.57}} \vspace{-1em}\caption{Output SINR
versus the number of iterations.} \label{fig:AVF_rank}
\end{minipage}
\end{figure}

In Fig. \ref{fig:AVF_rank}, we keep the same scenario as that in
Fig. \ref{fig:AVF_sve} (a) and check the iteration number for the
existing and proposed methods. The number of snapshots is fixed to
$N=500$. The most adequate iteration number for the proposed CCM-AVF
algorithm is $K=3$, which is comparatively lower than other analyzed
algorithms, but reach the preferable performance. We also checked
that this value is rather insensitive to the number of users in the
system, to the number of sensor elements, and work efficiently for
the studied scenarios.

\section{Concluding Remarks}
We developed an AVF algorithm based on the CCM design for robust
adaptive beamforming. The algorithm provides a positive measure of
the expected deviation of the beamformer output from the CM
condition and thus is robust against the severe scenarios. The
weight solution is iterated by jointly calculating the auxiliary
vector and the scalar factor, which iteratively exchange information
between each other and lead to an improved performance over prior
art. The selection of the iteration number may be more efficient and
adaptive with the change of the system (e.g., the number of users
change) if we employ other techniques \cite{Mathews}. We will
consider further improvements to the proposed CCM-AVF algorithm in
the near future.

\end{document}